\documentclass[sigconf]{acmart}
\AtBeginDocument{%
  }

\setcopyright{acmlicensed}
\copyrightyear{2018}
\acmYear{2018}
\acmDOI{XXXXXXX.XXXXXXX}
\acmConference[Conference acronym 'XX]{Make sure to enter the correct
  conference title from your rights confirmation email}{June 03--05,
  2018}{Woodstock, NY}
\acmISBN{978-1-4503-XXXX-X/2018/06}

\usepackage{amsmath}
\usepackage{multirow}
\usepackage{booktabs}
\usepackage{hyperref}
\usepackage{indentfirst}
\usepackage{subcaption}
\usepackage{xcolor}
\usepackage{svg}

\newcommand{\llmsforcode}{LLMs4Code\xspace}

\newcommand{\RQone}{To what extent do original and quantized LLMs generate
hallucinated packages in the shell commands?}
\newcommand{\RQtwo}{What proportion of the valid packages generated by LLMs
exhibit known security vulnerabilities?}
\newcommand{\RQthree}{What are the properties of the hallucinated packages of
the original and quantized \llmsforcode?}

\usepackage{seqsplit}
\newcommand{\ttt}[1]{%
\begingroup \protect
\renewcommand{\seqinsert}{\ifmmode\allowbreak\else\-\fi}%
\protect\texttt{\protect\seqinsert{\protect\seqsplit{\small#1}}}%
\endgroup }
\newcommand{\tttscript}[1]{%
\begingroup \protect
\renewcommand{\seqinsert}{\ifmmode\allowbreak\else\-\fi}%
\protect\texttt{\protect\seqinsert{\protect\seqsplit{\scriptsize#1}}}%
\endgroup }
\newcommand{\tttfoot}[1]{%
\begingroup \protect
\renewcommand{\seqinsert}{\ifmmode\allowbreak\else\-\fi}%
\protect\texttt{\protect\seqinsert{\protect\seqsplit{\footnotesize#1}}}%
\endgroup }

\usepackage{tcolorbox}
\newtcolorbox{boxK}{
    fontupper = \small,
    sharpish corners, 
    boxrule = 0pt,
    toprule = 0pt, 
}

\usepackage{xspace}
\begin{document}

\title{Secure or Suspect? Investigating Package Hallucinations of Shell Command in Original and Quantized LLMs}

  \author{Md Nazmul Haque}
  \affiliation{%
  \institution{North Carolina State University} \city{Raleigh, NC} \country{USA}}
  \email{mhaque4@ncsu.edu}

  \author{Elizabeth Lin}
  \affiliation{%
  \institution{North Carolina State University} \city{Raleigh, NC} \country{USA}}
  \email{etlin@ncsu.edu}

  \author{Lawrence Arkoh}
  \affiliation{%
  \institution{North Carolina State University} \city{Raleigh, NC} \country{USA}}
  \email{larkoh@ncsu.edu}

  \author{Biruk Tadesse}
  \affiliation{%
  \institution{North Carolina State University} \city{Raleigh, NC} \country{USA}}
  \email{batadess@ncsu.edu}

  \author{Bowen Xu}
  \affiliation{%
  \institution{North Carolina State University} \city{Raleigh, NC} \country{USA}}
  \email{bxu22@ncsu.edu}

\renewcommand{\shortauthors}{Haque et al.}

\begin{abstract}
Large Language Models for code (\llmsforcode) are increasingly used to generate software artifacts, including library and package recommendations in languages such as Go. However, recent evidence shows that LLMs frequently hallucinate package names or generate dependencies containing known security vulnerabilities, posing significant risks to developers and downstream software supply chains. At the same time, quantization has become a widely adopted technique to reduce inference cost and enable deployment of LLMs on resource-constrained environments. Despite its popularity, little is known about how quantization affects the correctness and security of LLM-generated software dependencies while generating shell commands for package installation.

In this work, we conduct the first systematic empirical study of the impact of quantization on package hallucination and vulnerability risks in LLM-generated Go packages. We evaluate five Qwen model sizes under full-precision, 8-bit, and 4-bit quantization across three datasets (SO, MBPP, and paraphrase). Our results show that quantization substantially increases the package hallucination rate (PHR), with 4-bit models exhibiting the most severe degradation. We further find that even among the correctly generated packages, the vulnerability presence rate (VPR) rises as precision decreases, indicating elevated security risk in lower-precision models. Finally, our analysis of hallucinated outputs reveals that most fabricated packages resemble realistic URL-based Go module paths, such as most commonly malformed or non-existent GitHub and golang.org repositories, highlighting a systematic pattern in how LLMs hallucinate dependencies. Overall, our findings provide actionable insights into the reliability and security implications of deploying quantized LLMs for code generation and dependency recommendation.

\end{abstract}



\received{2 December 2025}

\maketitle

 \section{Introduction}
 
In the software engineering domain, recent studies have shown that developers rely on Large Language Models (LLMs) as key assistants for automating a broad range of programming activities \cite{rahman2024code, fan2023large, liu2024exploring}.  These LLMs have become widely adopted in daily developer tasks such as code generation, summarization, bug fixing, and shell command generation\cite{rasnayaka2024empirical,meng2024empirical,ram2024shellcode}.
Within these activities, one increasingly common use case is \textit{shell command generation}. 
Developers frequently request LLMs to produce installation commands, environment setup scripts, 
and package management instructions to support rapid software development. These commands are 
particularly important when installing third-party packages, which help speed up 
development, reduce errors, and improve the overall quality of the resulting software systems.
However, despite their growing usefulness, modern LLMs come with substantial 
computational and memory requirements\cite{li2024efficient}. To address these limitations, lightweight 
variants—particularly \textit{quantized LLMs}—have been introduced. Quantization 
reduces the numerical precision of model weights\cite{lang2024comprehensive}, thereby lowering memory 
footprint and computational cost while still maintaining competitive performance 
across many coding tasks. Although these quantised models make LLMs more suitable 
for deployment in resource-constrained environments, both quantised and 
unquantised variants are still known to suffer from a persistent issue: 
\textit{hallucination}. This phenomenon occurs when an LLM 
produces outputs that are factually incorrect, irrelevant, or not grounded in the given prompt 
\cite{rahman2024code, fan2023large, liu2024exploring}. Hallucinations in software engineering 
contexts introduce serious risks, including misleading developers, introducing defects, and 
creating vulnerabilities in the development pipeline \cite{lee2025hallucination}. A critical 
effect of this issue arises when LLMs hallucinate package names or installation commands\cite{spracklen2025packageyoucomprehensiveanalysis}. 
Prior work has shown that such hallucinations can even lead to \textit{slopsquatting attacks}, 
where attackers register malicious packages in public registries using hallucinated names suggested by LLMs 
\cite{park2025slopsquatting}.

While hallucination is a known problem for full-precision LLMs, it raises an important question 
for their compressed counterparts. Lightweight LLM variants—particularly \textit{quantized LLMs}—
have been introduced to reduce memory footprint and computational requirements while still 
maintaining strong performance across many coding tasks. Several studies demonstrate that 
quantised and unquantised variants can achieve comparable results on tasks such as code 
generation and bug fixing. If quantised models also exhibit reduced hallucination when 
generating shell commands, this becomes highly valuable. In such cases, developers and model 
consumers may prefer quantised models over their original counterparts, not only for efficiency 
but also for improved reliability in security-sensitive tasks such as package installation. 

Given these considerations, this study seeks to investigate the rate of package-related 
hallucinations in the shell command outputs of quantised and unquantised variants of the same LLM.


Despite the potential impact of quantization on package hallucinations, its effect has not been empirically validated. Hence, we propose our central research question (RQ):
Thus, we propose our central research question as follows:
\begin{tcolorbox}
\textit{How does quantization impact package hallucination while generating shell commands?}
\end{tcolorbox}

To investigate this RQ, we first curated three distinct datasets capturing real-world shell command generation scenarios. We then applied a state-of-the-art quantization technique to the leading open-source \llmsforcode model, \emph{Qwen}, across multiple parameter scales. After quantization, we conducted a systematic evaluation to measure its effect on package hallucination, vulnerable package recommendation, and overall command accuracy.

Our main contributions are summarized as follows:

\begin{itemize}
    \item We curated and released three datasets tailored for evaluating package hallucination in shell command generation.
    \item We applied state-of-the-art quantization techniques to Qwen models of various sizes and conducted a comprehensive assessment of their performance.
    \item We propose the first empirical analysis of how quantization affects package hallucination in command-generation tasks.
    \item We provide guidance for practitioners on balancing model size, quantization level, and reliability when deploying LLMs for shell automation.
\end{itemize}

\noindent\textbf{Data availability}.
We release experimental results and replication package at \url{https://anonymous.4open.science/r/phrGo-39EB}.
  \section{Threat Model}
\label{sec:threat-model}

\begin{figure*}
    \centering
    \includegraphics[width=1.5\columnwidth]{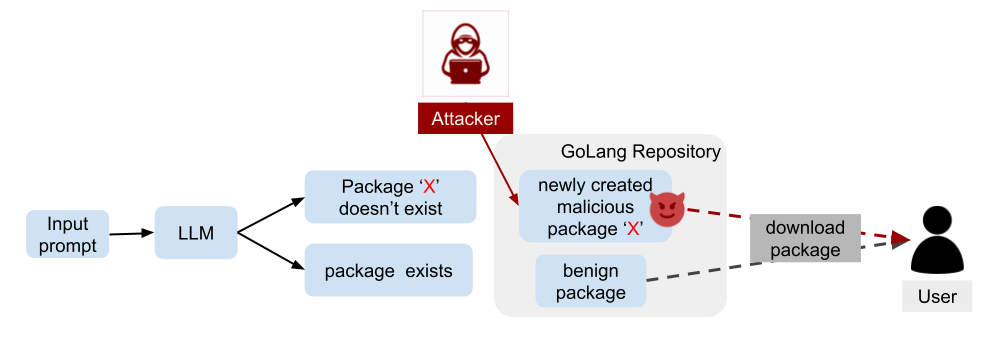}
    \caption{Threat Model}
    \label{fig:threat-model}
\end{figure*}

In this section, we discuss our threat model, as shown in Figure~\ref{fig:threat-model}.
In the threat model, there are two parties: 1) A user who wants to install packages and 2) An attacker creating malicious packages.
There is another entity: LLM, that accepts user input and generates responses.
The user prompts the LLM for shell commands to install golang packages, here we assume the user blindly trusts the LLM responses and executes whatever response is given by the LLM.
As the user blindly trusts the LLM, it could be executing commands that include hallucinated packages.

In our threat model, the attacker can take advantage of the hallucinated packages.
The attacker could observe what packages are frequently hallucinated by the LLM and create a new malicious golang package that matches the hallucinated string.
This allows to attacker to distribute malicious scripts through golang packages.
If the user executes the command from the LLM after the attacker creates the malicious package, they will unintentionally download a malicious package.

  \section{Related Work}
Large Language Models (LLMs) have been widely adopted in software engineering (SE) tasks, including code generation~\cite{chen2021evaluatinglargelanguagemodels, rozière2024codellamaopenfoundation}, bug fixing~\cite{10359304, 10172803}, code summarization~\cite{ahmad-etal-2020-transformer, wang-etal-2023-codet5}, and API or library recommendation~\cite{latendresse2024chatgptgoodsoftwarelibrarian, 10.1145/3276517}. In many of these tasks, the generated code relies on external packages or dependencies, which developers must install before the code executes. Prior work shows that LLM outputs often influence which dependencies developers choose~\cite{mohsin2024trustlargelanguagemodels}, and incorrect package suggestions can introduce security issues that spread through the software supply chain.

Although there is extensive research on LLM-generated source code, \textbf{no existing work explicitly studies shell command generation for Go.} Existing studies focus primarily on the correctness of code, syntax errors, test outcomes, or import statements, rather than on the shell commands that install and execute external modules. This is a significant gap because shell commands can directly modify system environments, fetch remote code, or install packages with elevated privileges. As a result, mistakes in shell command generation carry more direct and severe consequences than errors in the source code itself.

Recent studies have identified \emph{package hallucination} as a new category of LLM output failures, where the model suggests packages that do not exist in official registries. Spracklen et al.\ conducted the first large-scale study of this phenomenon across Python and JavaScript ecosystems~\cite{spracklen2025packageyoucomprehensiveanalysis}. They analyzed 16 commercial and open-source LLMs and found that almost \(20\%\) of the packages suggested by these models were hallucinated, and open-source models hallucinated four times more often than commercial ones. Their results also showed that these hallucinations are not random; models tend to repeat the same fake package names multiple times, which makes it easier for malicious actors to register those names and deceive developers.

At the same time, quantization has become a common model compression technique that reduces memory usage and inference cost while maintaining acceptable accuracy~\cite{NEURIPS2022_c3ba4962, frantar2023gptqaccurateposttrainingquantization}. Quantized LLMs are increasingly deployed on edge devices, IDE plugins, and local developer setups. While prior work has examined how quantization affects code generation quality~\cite{NEURIPS2023_1feb8787}, no study has explored its effect on hallucination behavior, especially in the context of shell command generation or dependency installation.

In contrast to existing work, we investigate hallucinations in \textbf{Go shell commands} generated by both original and quantized variants of the same LLMs. We study whether reduced numerical precision amplifies hallucination rates and whether hallucinated packages carry known security vulnerabilities. To the best of our knowledge, \emph{this is the first study} to analyze hallucination behavior in shell command generation, the Go ecosystem, and quantized LLMs simultaneously.

\label{sec:relatedWork}
  \section{Research Question}
\label{sec:researchQuestion} To guide our investigation, we formulate the following
research questions (RQs) and corresponding motivations, focusing on evaluating
the relationship between model quantization and shell command hallucination in
\llmsforcode.

\textbf{RQ1:} \RQone \label{RQ1}

\textbf{Motivation:} Quantization is widely adopted to reduce the computational and
memory footprint of \llmsforcode, enabling their deployment on resource-constrained
devices. However, the impact of quantization on the accuracy and reliability of
shell command generation remains underexplored. Quantization changes numerical precision
and internal representations, which can shift token probabilities and
calibration. For generation tasks that produce exact, security-sensitive
artifacts like shell commands, even small distributional shifts can turn plausible
outputs into incorrect or dangerous commands (e.g., destructive file operations
or privilege escalations). This research question aims to investigate how
different levels of quantization influence the rate of shell command
hallucination. Understanding this relationship is crucial for developers and researchers
aiming to deploy efficient yet reliable \llmsforcode in real-world applications.

\textbf{RQ2:} \RQtwo \label{RQ2}

\textbf{Motivation:} Packages recommended by \llmsforcode may introduce known
vulnerabilities into software projects that directly affect dependency hygiene
and supply-chain risk. Vulnerable dependencies can be exploited to execute arbitrary
code, escalate privileges, or exfiltrate data, and even a single insecure
package—particularly when pulled in transitively—can compromise an entire
application. Moreover, LLMs may suggest outdated, unmaintained, or omitted
package versions that contain well-documented CVEs. Quantifying the proportion of
generated packages that are known-vulnerable (and characterizing their severity
and exploitability) will reveal the practical risk of adopting model-suggested dependencies
in production. These findings can guide mitigations, such as automated vulnerability
scanning, stricter dependency pinning, or model-side filters, and help
practitioners balance the convenience of model assistance against real-world
security requirements.

\textbf{RQ3:} \RQthree \label{RQ3}

\textbf{Motivation:} While vulnerability counts provide a baseline security metric,
examining the qualitative aspects of hallucinated packages reveals deeper
insights into model behavior. Key characteristics worth investigating include naming
patterns that may indicate plausibility, whether packages actually exist in
registries, how version formats are structured, and signals related to package
popularity in the cross-language ecosystem. This characterization serves multiple
purposes. First, it helps explain why models produce certain types of hallucinations,
whether they tend toward improbable names, malformed version strings, or outputs
resembling typosquatting attacks. Second, understanding how quantization affects
these patterns can reveal whether compression techniques introduce systematic
biases in package suggestions. Finally, these insights enable practical
applications such as automated detection systems that flag suspicious package
recommendations and prompt engineering strategies that reduce hallucination
rates.
  \section{Study Design}
\label{sec:studyDesign}
\begin{figure*}
    \centering
    \includegraphics[width=\linewidth]{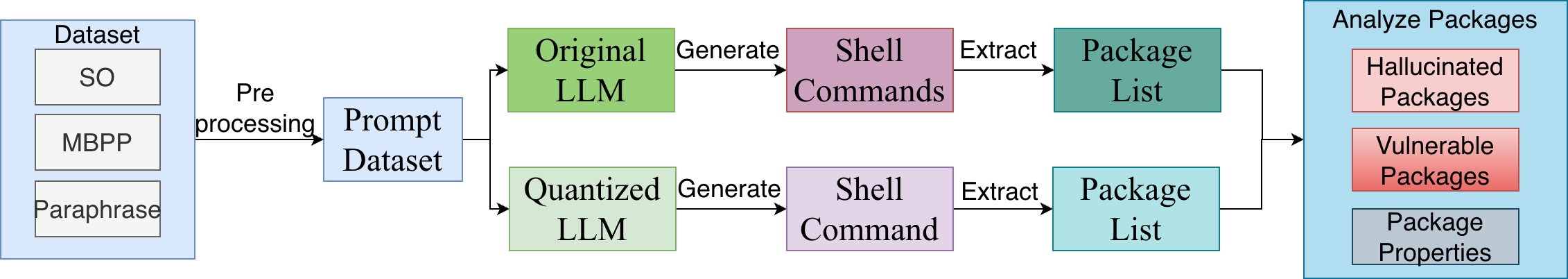}
    \caption{Overview of our study design}
    \label{fig:study_design}
\end{figure*}
We provide an overview of our study design in \autoref{fig:study_design} to address our core research question, as described in \autoref{sec:researchQuestion}. Our methodology involves assessing original and quantized models, analyzing their impact on shell command generation.
\subsection{Dataset Description}
\label{subsec:dataset_description}

There is no existing dataset that contains LLM-generated software packages along
with annotations regarding their hallucination status and vulnerability. Therefore,
we constructed three datasets from the following different sources for this
study.
\begin{itemize}
    \item \textbf{Stack Overflow (SO)}: We extract programming-related questions
        from Stack Overflow (SO) that are tagged with "golang" and like "golang".
        We use the Stack Exchange Data Explorer \footnote{https://data.stackexchange.com/}
        to query and retrieve these questions. Intially, we got 10,157 SO
        questions. Then, we use gpt-4o with zero shot to generate generic task. Many
        of these SO questions do not require any packages to install. From this
        motivation, we again use gpt-4o as judge to generate shell commands.
        Finally, we filter out the questions that do not generate any package installation
        commands. After this filtering step, we obtain a total of 5,435 SO questions
        that are suitable for generating GoLang packages.

    \item \textbf{MBPP}: We utilize the "text-to-code" generation prompts from the
        MBPP dataset \cite{austin2021mbpp} to generate software packages. These
        prompts are specifically designed to test the code generation
        capabilities of LLMs for the python language. However, we adapt these
        prompts to generate generic prompts in the preprocessing step which can
        be used for any programming language, including GoLang. After the
        preprocessing step, we obtain 974 prompts that are suitable for generating
        GoLang packages.

    \item \textbf{Paraphrase}: As our main goal is to evaluate the hallucination
        and vulnerability of LLM-generated packages, we need to ensure that the
        prompts are diverse enough to cover a wide range of scenarios. To achieve
        this, we use \emph{awesome-go} \footnote{https://github.com/avelino/awesome-go}
        as a source of GoLang packages. We extract the package names and
        corresponding metadata from the \emph{awesome-go} repository and feed them
        to gpt-4o with zero shot instruction to create paraphrased prompts. This
        results in a total of 1,746 unique prompts that are used to generate
        GoLang packages.
\end{itemize}

\subsection{Models Selection}
\label{subsec:models_selection}

For our study, we selected Qwen \llmsforcode from the EvalPlus leaderboard
\cite{liu2023your} as it is open-source, has different parameters, can be run
locally, and are suitable for code generation tasks. Qwen \llmsforcode is a
family of models that includes Qwen-0.5B, Qwen-1.5B, Qwen-3B, Qwen-7B, and Qwen-14B.

\subsection{Package Hallucination Detection}
\label{subsec:packages_detection} Hallucinated packages are packages that do not
exist in package repositories. External golang modules are installed through \ttt{go~get~ [url]}
or \ttt{go~install~[url]}. The \ttt{go} command will search up the \emph{url}
specified through the \ttt{GOPROXY}, if the package at the url is found, it is downloaded.
In this project, we checked hallucination by checking if the \emph{url} in the
go command is reachable through network requests. If a url is not reachable, we
determine the package as hallucinated.

\subsection{Evaluation Metrics}
\label{subsec:evaluation_metrics} We consider the set of package references
generated by an LLM when it produces shell commands. Let $G$ denote the set of all
generated packages. We partition $G$ into the following subsets:

\begin{itemize}
    \item $H \subseteq G$ (\emph{Hallucinated Packages}): The package names that
        do not correspond to any of the GoLang public registries or repositories.

    \item $E \subseteq G$ (\emph{Existing/Valid Packages}): the generated packages
        that \emph{exist} in public registries.

    \item $V \subseteq E$ (\emph{Vulnerable Packages}): the subset of existing packages
        that are flagged as vulnerable by a vulnerability oracle. We use OSV's
        API \footnote{https://osv.dev/\#use-the-api} as a vulnerability oracle
        in this work.
\end{itemize}
\begin{figure}
    \centering
    \includegraphics[width=\linewidth]{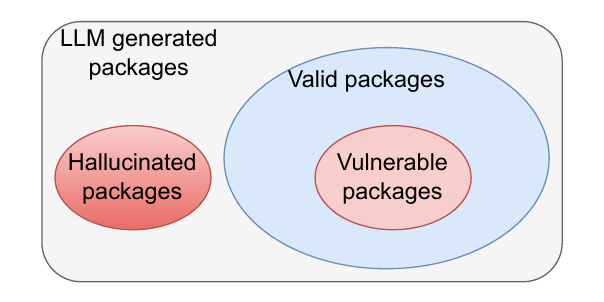}
    \caption{Categorization of LLM-generated output}
    \label{fig:llm_output}
\end{figure}
To make it better understood, we present it in \autoref{fig:llm_output}.
To address our research questions, we use two performance metrics:
\begin{itemize}
    \item \textbf{Package Hallucination Rate (PHR):} This metric addresses RQ1
        by quantifying the extent to which the LLM generates non-existent
        packages, defined in Equation \eqref{eq:phr}.
        \begin{align}
            \mathrm{PHR} & = \frac{|H|}{|G|}\label{eq:phr}
        \end{align}
        where $|H|$ is the number of hallucinated packages and $|G|$ is the total
        number of generated packages.

    \item \textbf{Vulnerable Package Rate (VPR):} This metric addresses RQ2 by
        measuring how often the LLM suggests existing packages that are known to
        be vulnerable, defined in \eqref{eq:vpr}.
        \begin{align}
            \mathrm{VPR} & = \frac{|V|}{|E|}\label{eq:vpr}
        \end{align}
        where $|V|$ is the number of vulnerable existing packages, and $|E|$ is
        the number of generated packages that match real packages.
\end{itemize}
Both metrics are bounded in $[0,1]$ and are defined when $|G|>0$ and $|E|>0$,
respectively.

\subsection{Implementation Details}
\label{subsec:implementation_details}

For qwen model, we used the model from the Hugging Face Hub \footnote{https://huggingface.co}.
We use the \texttt{transformers} library to load and run these models. For quantization,
we used bitsAndBytes, which supports 8- and 4-bit quantization. The models are run
on a machine with an NVIDIA A100 GPU with 80GB of memory.
  \section{Results}
\label{sec:results}

In this section, we discuss the results from our experiments to address the
research questions listed in \autoref{sec:researchQuestion}.

\begin{figure*}[ht]
    \centering
    \begin{subfigure}
        [b]{0.33\textwidth}
        \centering
        \includegraphics[width=\textwidth]{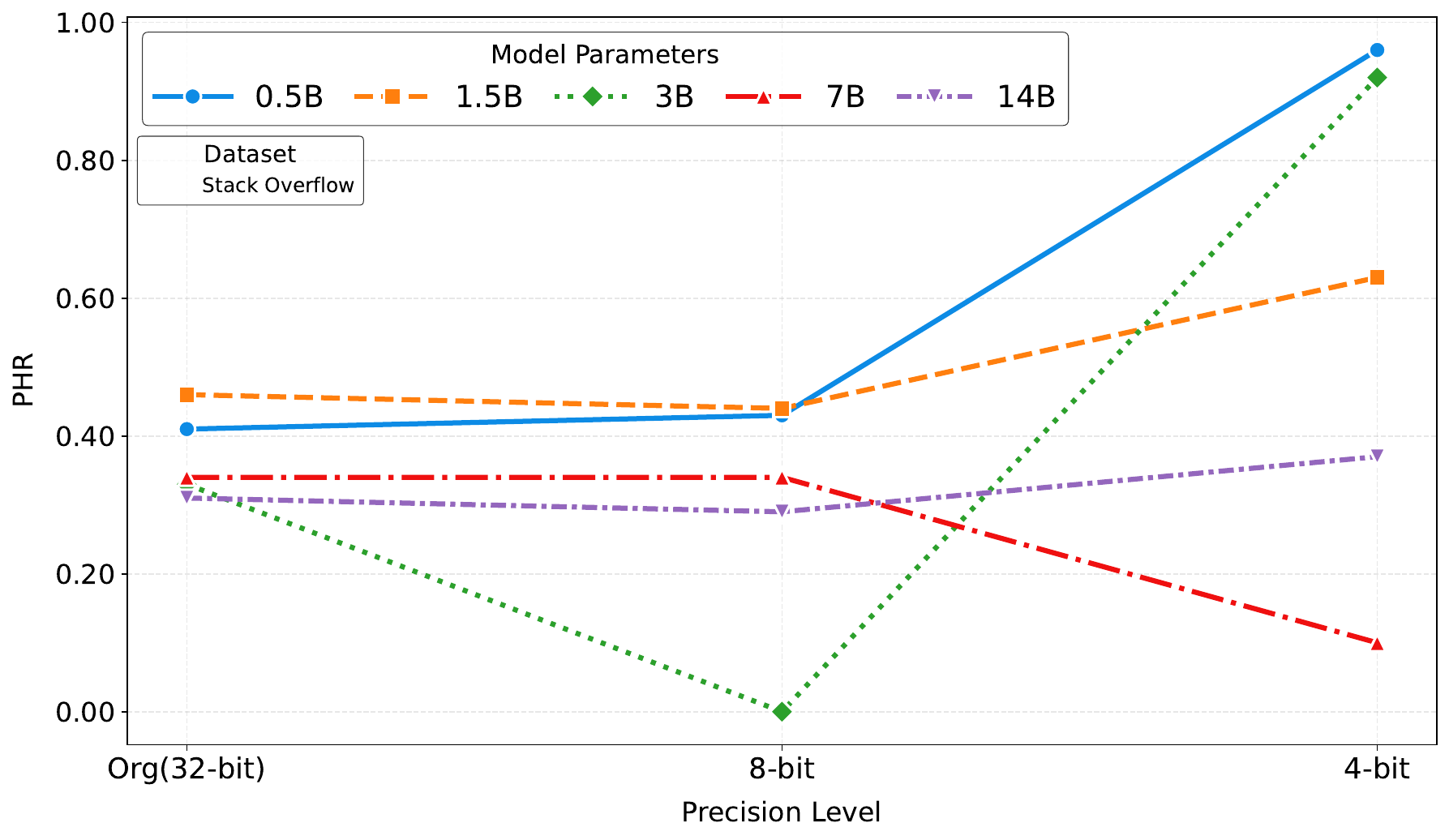}
        \caption{SO}
    \end{subfigure}
    \hfill
    \begin{subfigure}
        [b]{0.33\textwidth}
        \centering
        \includegraphics[width=\textwidth]{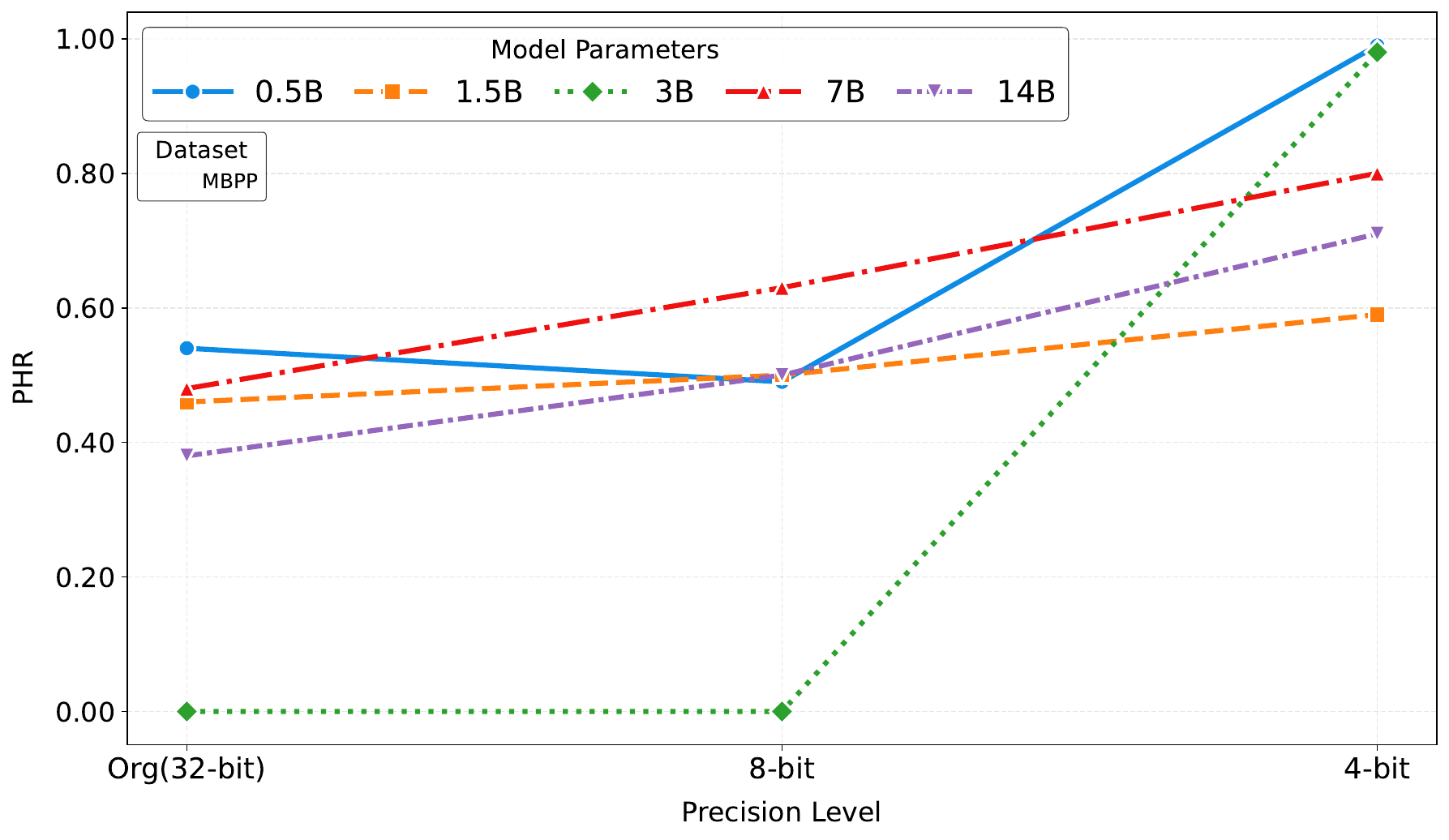}
        \caption{MBPP}
    \end{subfigure}
    \hfill
    \begin{subfigure}
        [b]{0.33\textwidth}
        \centering
        \includegraphics[width=\textwidth]{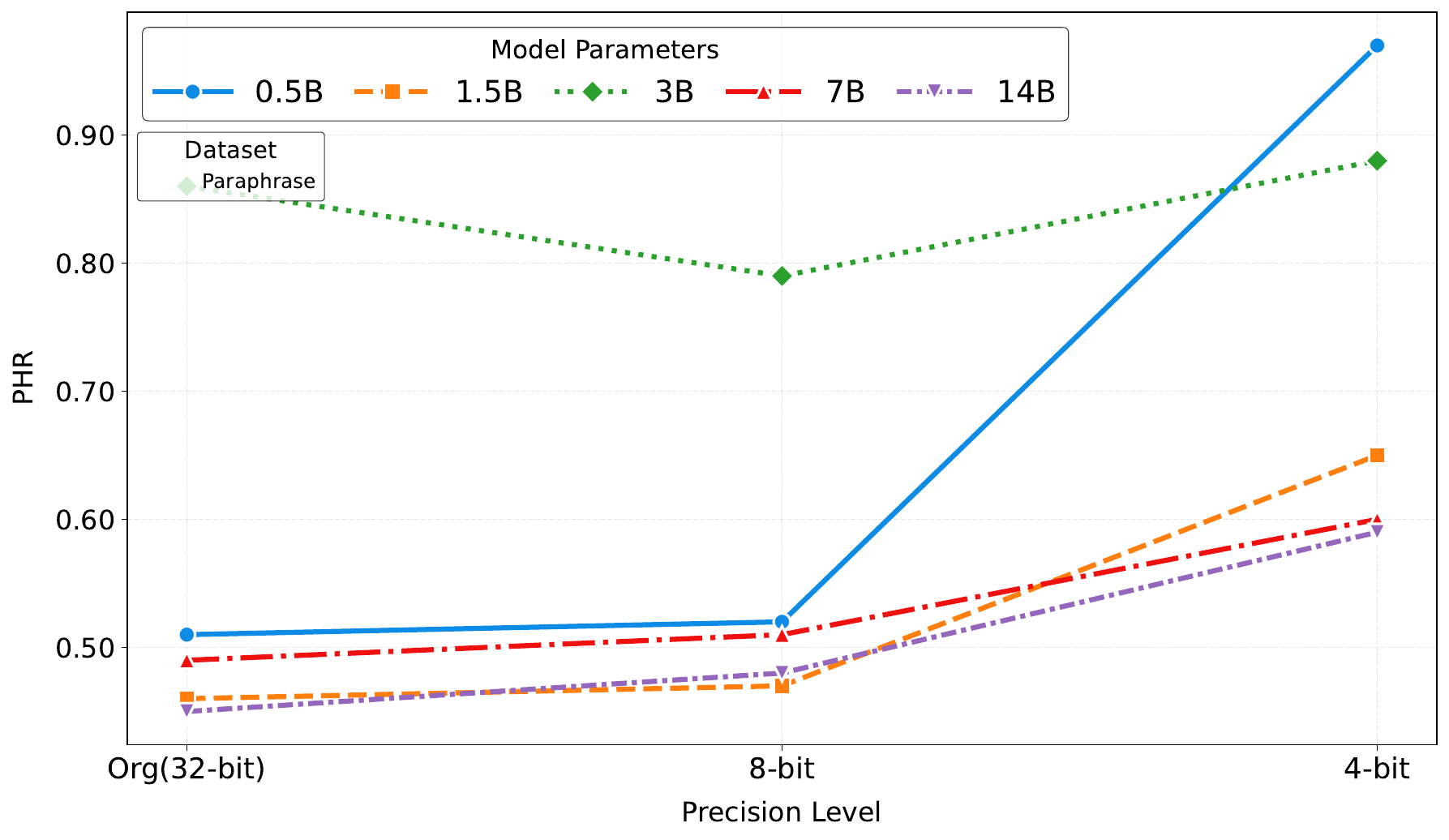}
        \caption{Paraphrase}
    \end{subfigure}
    \caption{Comparison of \textbf{PHR} over different precision levels of quantization.
    Lines represent different Qwen models, distinguished by parameter sizes.}
    \label{fig:RQ1}
\end{figure*}
\subsection{RQ1: Impact of Quantization on Package Hallucination}
To answer \emph{RQ1}, we evaluated the package hallucination rates (PHRs) of models across different quantization levels. The detailed results are shown in \autoref{fig:RQ1}, with exact values provided in \autoref{table:evaluation-results}.

When comparing the original full-precision models to their 8-bit quantized counterparts, we observe a moderate and consistent increase in package hallucination rates. Across all model sizes and datasets, the PHR typically rises by a few percentage points. For instance, in the 0.5B model, the PHR on the Stack Overflow dataset increases from 40.60\% (Full) to 43.04\% (8-bit). This trend indicates that even mild quantization begins to erode the model's precision in generating correct package names, though the degradation remains manageable.

In contrast, the transition from full precision to 4-bit quantization results in a dramatic and often catastrophic surge in hallucination. For smaller models, this effect is extreme, such as the PHR for the 0.5B model on the MBPP dataset, which increases from 53.58\% to 98.78\%. Larger models, while more robust, still suffer significant degradation, with the notable exception of the 7B model on the SO dataset, where PHR surprisingly improved from 34.00\% to 10.21\%. This severe impact demonstrates that aggressive 4-bit quantization critically compromises a model's knowledge recall, leading to a proliferation of non-existent package names in its outputs.

Overall, quantization systematically increases package hallucination, with the severity of the effect directly correlated to the aggressiveness of the compression. The 8-bit quantization introduces a measurable but often acceptable decline in accuracy, while 4-bit quantization poses a fundamental threat to functional reliability, rendering smaller models nearly unusable for accurate package generation.

\begin{boxK}
    \textit{\textbf{Finding:}} Quantization consistently increases package hallucination across model sizes and datasets. Lower precision leads to higher hallucination rates, with 4-bit quantized models showing the most severe degradation in reliability.
\end{boxK}

\begin{table*}[t]
    \centering
    \scriptsize
    \caption{Evaluation results of LLM generated packages}
    \label{table:evaluation-results}
    \begin{tabular}{cc|ccc|ccc|ccc}
        \toprule
        
        \multirow{2}{*}{Parameters} & \multirow{2}{*}{Precision} & \multicolumn{3}{c}{\textbf{Stack Overflow}} & \multicolumn{3}{c}{\textbf{MBPP}} & \multicolumn{3}{c}{\textbf{Paraphrase}} \\

        & & go pkgs & PHR & VPR & go pkgs & PHR & VPR & go pkgs & PHR & VPR \\

        \midrule

        \multirow{3}{*}{0.5B} & Full & 8748 & 40.60\% & 12.49\% & 642 & 53.58\% & 8.39\% & 2795 & 50.84\% & 15.57\% \\
        &  8bit & 8652 & 43.04\% & 12.24\% & 636 & 49.21\% & 8.36\% & 2872 & 52.40\% & 17.63\% \\
        & 4-bit & 8431 & 96.11\% & 10.98\% & 658 & 98.78\% & 12.50\% & 2185 & 96.66\% & 8.22\% \\

        \hline

        \multirow{3}{*}{1.5B} & Full & 21439 & 46.17\% & 13.70\% & 1303 & 46.43\% & 4.73\% & 6150 & 45.79\% & 13.44\% \\
        & 8bit & 19566 & 44.16\% & 13.51\% & 1077 & 49.77\% & 3.51\% & 5803 & 47.29\% & 11.02\% \\
        & 4-bit & 7550 & 62.93\% & 15.83\% & 1084 & 58.76\% & 9.17\% & 4700 & 64.60\% & 10.76\% \\

        \hline

        \multirow{3}{*}{3B} & Full & 3 & 33.33\% & 0.00\% & 0 & 0.00\% & 0.00\% & 28 & 85.71\% & 0.00\% \\
        & 8bit & 1 & 0\% & 0.00\% & 0 & 0.00\% & 0.00\% & 90 & 78.89\% & 5.26\% \\
        & 4-bit & 1568 & 92.09\% & 11.29\% & 125 & 97.60\% & 0.00\% & 734 & 87.74\% & 11.11\% \\

        \hline

        \multirow{3}{*}{7B} & Full & 4185 & 34.00\% & 12.89\% & 155 & 48.39\% & 1.25\% & 1864 & 49.41\% & 10.92\% \\
        & 8bit & 4037 & 34.06\% & 12.40\% & 147 & 69.59\% & 0.00\% & 1777 & 50.76\% & 10.17\% \\
        & 4-bit & 1655 & 10.21\% & 19.58\% & 126 & 80.16\% & 0.00\% & 1692 & 60.11\% & 13.63\% \\

        \hline

        \multirow{3}{*}{14B} & Full & 403 & 30.52\% & 16.43\% & 13 & 38.46\% & 0.00\% & 1132 & 45.41\% & 12.30\% \\
        & 8bit & 537 & 29.24\% & 13.68\% & 8 & 50.00\% & 0.00\% & 1100 & 48.27\% & 12.48\% \\
        & 4-bit & 682 & 37.10\% & 14.92\% & 7 & 71.43\% & 0.00\% & 977 & 59.37\% & 9.57\% \\
        
        \bottomrule

    \end{tabular}

\end{table*}

\begin{figure*}[ht]
    \centering
    \begin{subfigure}
        [b]{0.33\textwidth}
        \centering
        \includegraphics[width=\textwidth]{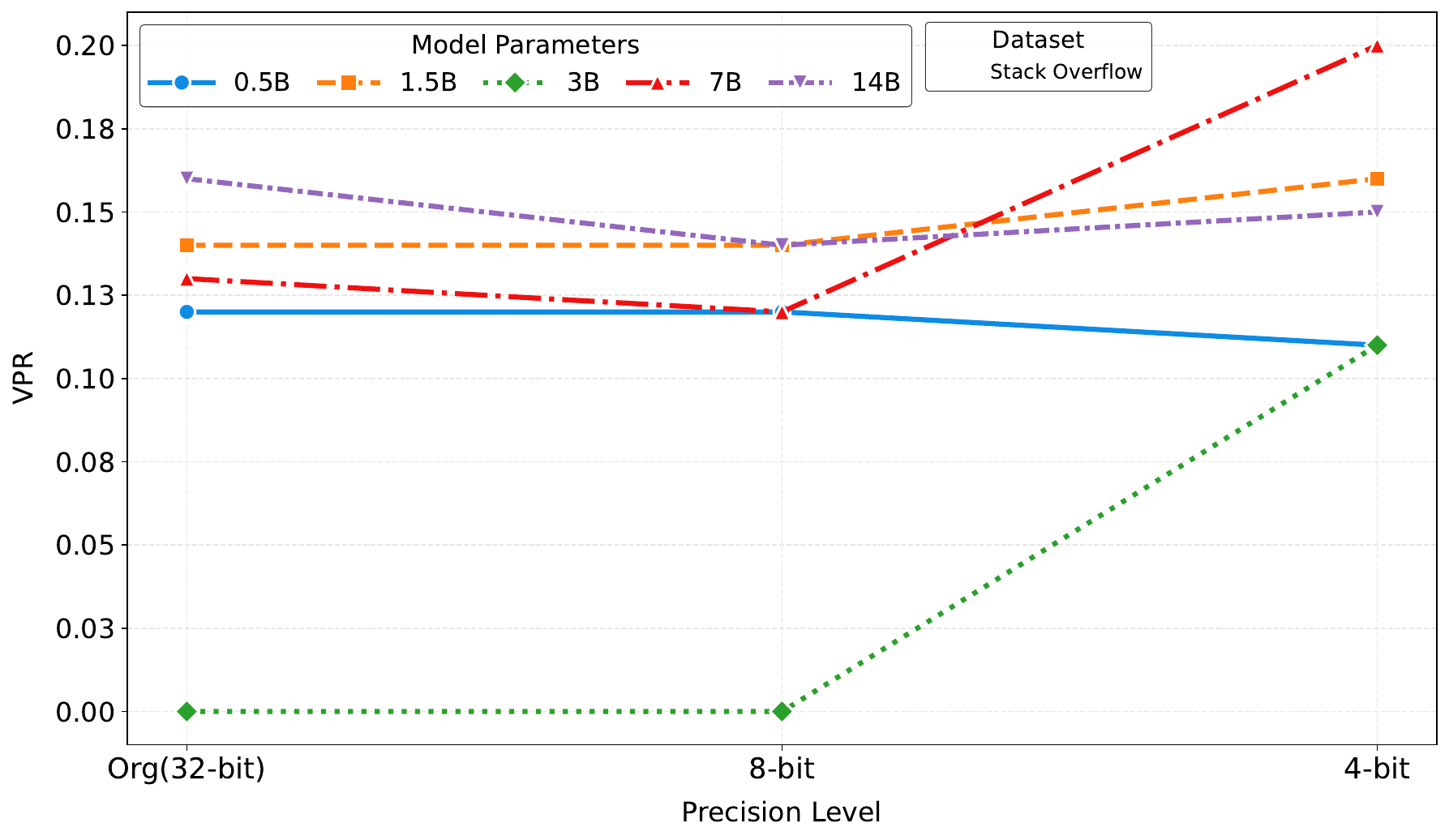}
        \caption{SO}
    \end{subfigure}
    \hfill
    \begin{subfigure}
        [b]{0.33\textwidth}
        \centering
        \includegraphics[width=\textwidth]{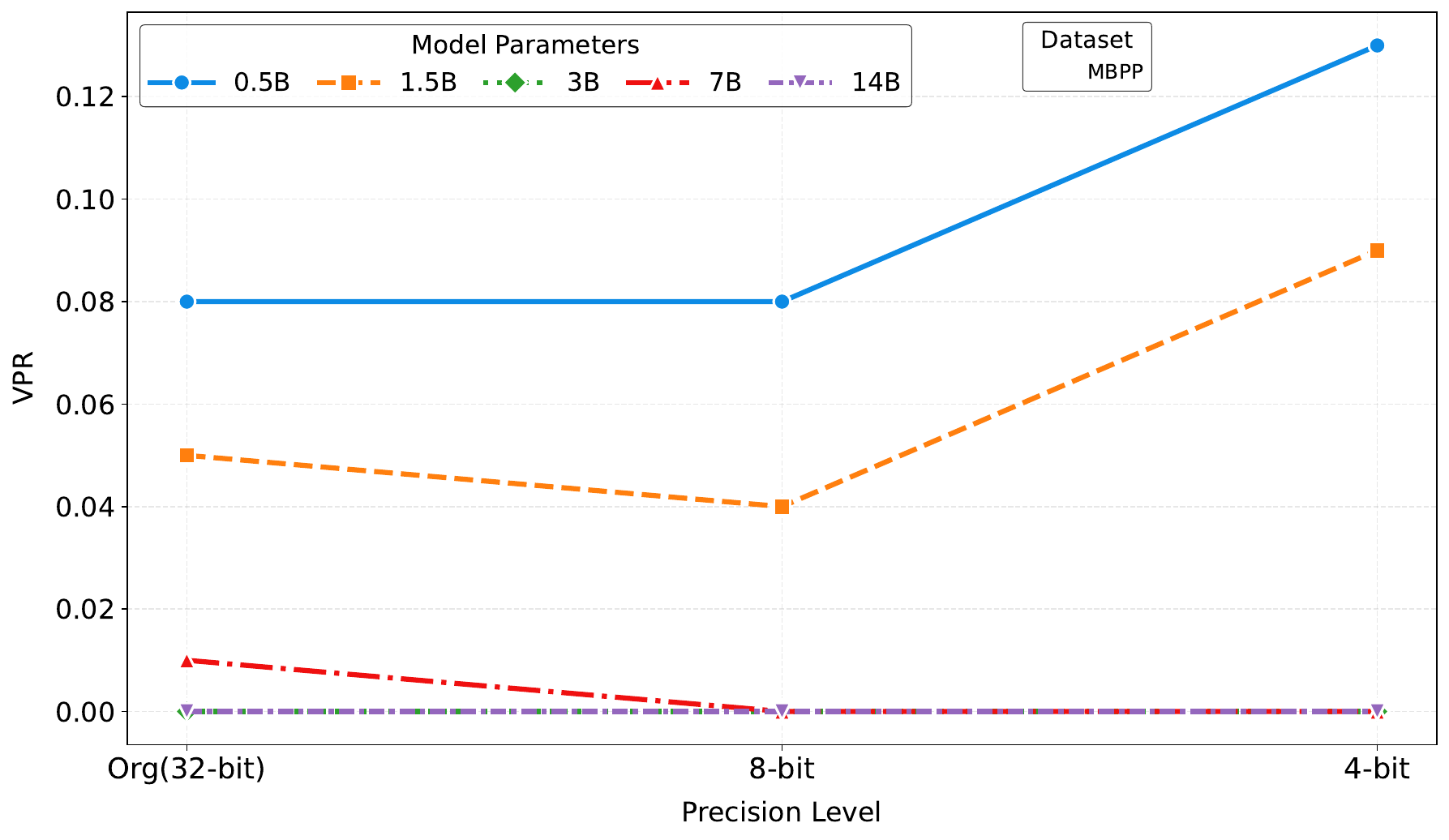}
        \caption{MBPP}
    \end{subfigure}
    \hfill
    \begin{subfigure}
        [b]{0.33\textwidth}
        \centering
        \includegraphics[width=\textwidth]{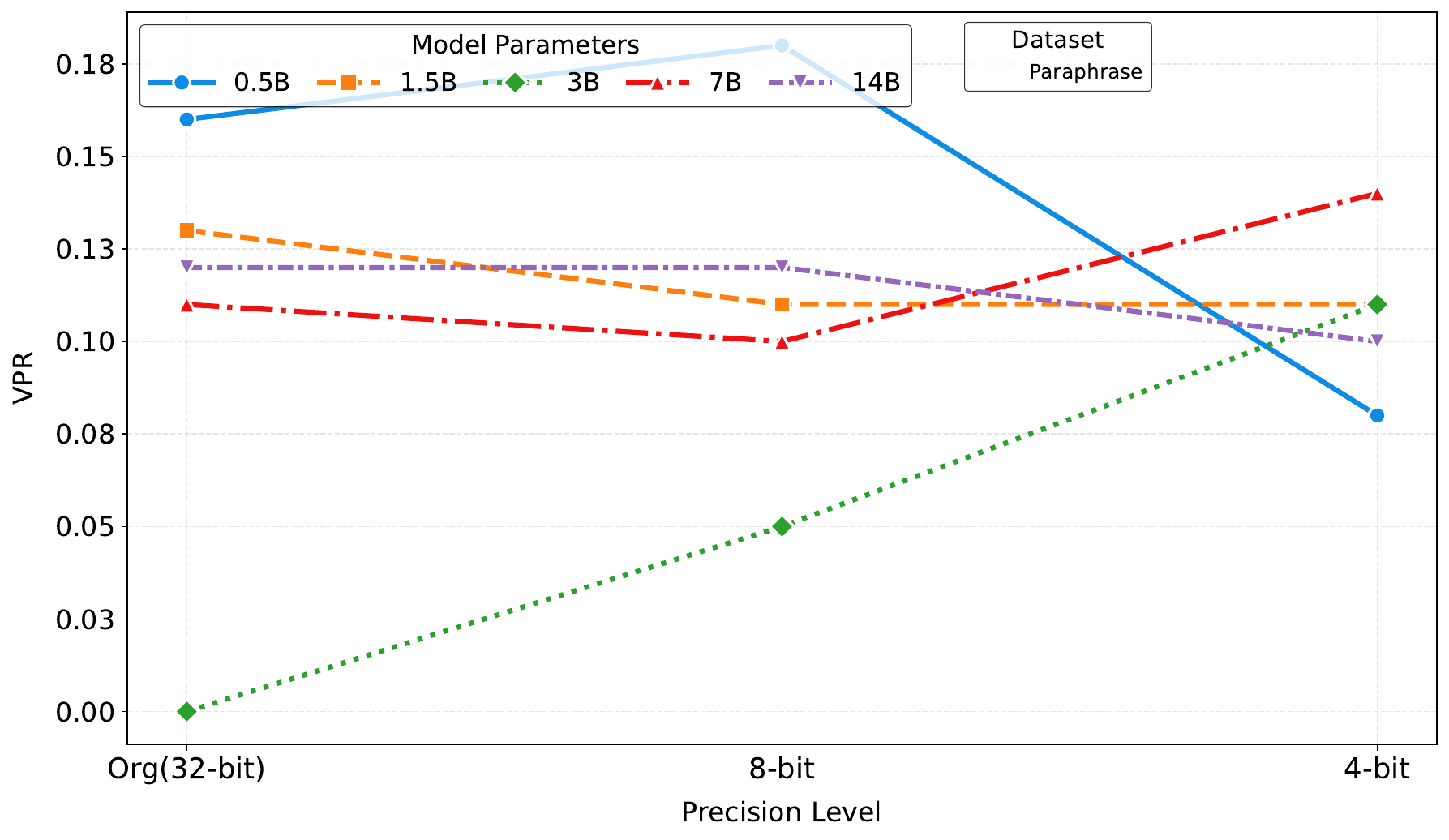}
        \caption{Paraphrase}
    \end{subfigure}
    \caption{Comparison of \textbf{VPR} over different precision levels of quantization.
    Lines represent different Qwen models, distinguished by parameter sizes.}
    \label{fig:RQ2}
\end{figure*}

\subsection{RQ2: Vulnerabilities in LLM Generated Packages}
To answer \emph{RQ2}, we analyzed the presence of known vulnerabilities in the *valid/existing* packages generated by quantized models, calculating the vulnerability presence rate (VPR) as detailed in Section~\ref{sec:studyDesign}. Results are shown in \autoref{fig:RQ2} and \autoref{table:evaluation-results}.

When quantizing from full precision to 8-bit, the impact on vulnerability rates varies significantly by model size. For smaller models (0.5B, 1.5B), the VPR typically shows minimal change or a slight decrease. For instance, the VPR for the 0.5B model on the SO dataset shifts marginally from 12.49\% to 12.24\%. In contrast, larger models (7B, 14B) often exhibit a clearer reduction in VPR; the 7B model's VPR on the paraphrase dataset drops from 10.92\% to 10.17\%. This suggests that 8-bit quantization in larger models might act as a mild filter, slightly reducing the propensity to suggest vulnerable packages among its valid outputs, whereas smaller models show inconsistent or negligible effects.

The effect of 4-bit quantization on vulnerability rates is more severe and reveals a stark model-size dependency. For smaller models like 0.5B and 1.5B, the VPR among the few remaining valid packages shows considerable fluctuation, sometimes decreasing (e.g., 0.5B on SO: 12.49\% to 10.98\%) and other times increasing sharply (e.g., 1.5B on SO: 13.70\% to 15.83\%). For larger models, the trend is more alarming: the 7B model's VPR on the SO dataset surges from 12.89\% to 19.58\%. This indicates that aggressive 4-bit compression catastrophically distorts the "knowledge" of larger models, corrupting their package recall in a way that disproportionately surfaces vulnerable dependencies. The few packages these models still generate correctly are substantially riskier.

In summary, quantization's impact on code security is dual-faceted and scale-dependent. The 8-bit compression shows a neutral to slightly positive effect, particularly in larger models, while 4-bit quantization introduces a critical security threat by significantly inflating the vulnerability rate within the already scarce set of valid packages, an effect most pronounced in larger-scale models. This establishes that aggressive model compression for efficiency can inadvertently and severely compromise the security of AI-generated code.

\begin{boxK}
    \textit{\textbf{Finding:}} Quantization correlates with an increased vulnerability presence rate in valid packages. While the primary effect is more frequent hallucination, the packages that are correctly generated by lower-precision models also tend to be more vulnerable.
\end{boxK}

\subsection{RQ3: Properties in LLM Hallucinated Packages}
Our third research question is: \emph{\RQthree} 
To understand the properties, we observed the hallucinated package strings. 
The numbers from all models and quantization methods would create an enormous table, thus we selected the hallucination distribution of one of the models (1.5B 4-bit quantization) as Table~\ref{table:hallucination-distribution}.
As shown in Table~\ref{table:hallucination-distribution}, we categorized
the hallucinated package strings into \emph{url}, \emph{filepath}, and \emph{other}. 
As most of the golang packages are hosted on \emph{github.com} and \emph{golang.org}, we further separated those two as their own category in \emph{url}.

Across all models, the LLM generated package strings mostly in the \emph{url} category, with the majority being \emph{github.com} and \emph{golang.org} urls.
We observe that more than half of the generated \emph{github.com} urls are hallucinated.
Similarly, the majority of the \emph{other} urls are also hallucinated.
The large number of hallucinated github urls shows that the LLM can correctly understand that golang packages are often hosted on github, however, it can hallucinate repository names.
For the \emph{other} url strings, we found the strings typically are in existing domains that host go packages, such as \emph{gopkg.in}.
However, as is the case with github repositories, the LLM struggles to correctly generate the subpaths for the url.

Finally, in the table, we identified a string in the \emph{other} category that exists.
We inspected this and the string was \ttt{your.package.name}.
This was categorized as \emph{other} category as it is unclear if this is a url.
However, the string is actually a url that exists, showing how strings that are seemingly benign could also results in unintended consequences when executed in shell commands.

\begin{boxK}
    \textit{\textbf{Finding:}} LLMs can hallucinate golang package urls. With hallucinated Github repositories being the largest attack vector. 
\end{boxK}

\begin{table*}[t]
    \centering
    \scriptsize
    \caption{Hallucination distribution of LLM generated packages for 1.5B 4bit quantization}
    \label{table:hallucination-distribution}
    \begin{tabular}{cc|cc|cc|cc}
        \toprule

        \multicolumn{2}{c|}{\multirow{2}{*}{\textbf{String category}}} & \multicolumn{2}{c|}{\textbf{Stack Overflow}} & \multicolumn{2}{c|}{\textbf{MBPP}} & \multicolumn{2}{c}{\textbf{Paraphrase}} \\

        & & exists & hallucinated & exists & hallucinated & exists & hallucinated \\

        \midrule

        \multirow{3}{*}{url} & github.com  & 1794 & 1868 & 98 & 447 & 1126 & 1861\\
        &  golang.org & 721 & 42 & 313 & 20 & 401 & 29\\
        & other & 283 & 508 & 36 & 58 & 137 & 664\\

        \hline

        \multicolumn{2}{c|}{filepath} & 0 & 1372 & 0 &37 & 0 & 217\\

        \hline

        \multicolumn{2}{c|}{other} & 1 & 961 & 0 & 75 & 0 & 265 \\
        
        \bottomrule

    \end{tabular}

\end{table*}
  \section{Discussion}
\label{sec:discussion}
\begin{figure*}[ht]
    \centering
    \begin{subfigure}
        [b]{0.49\textwidth}
        \centering
        \includegraphics[width=\textwidth]{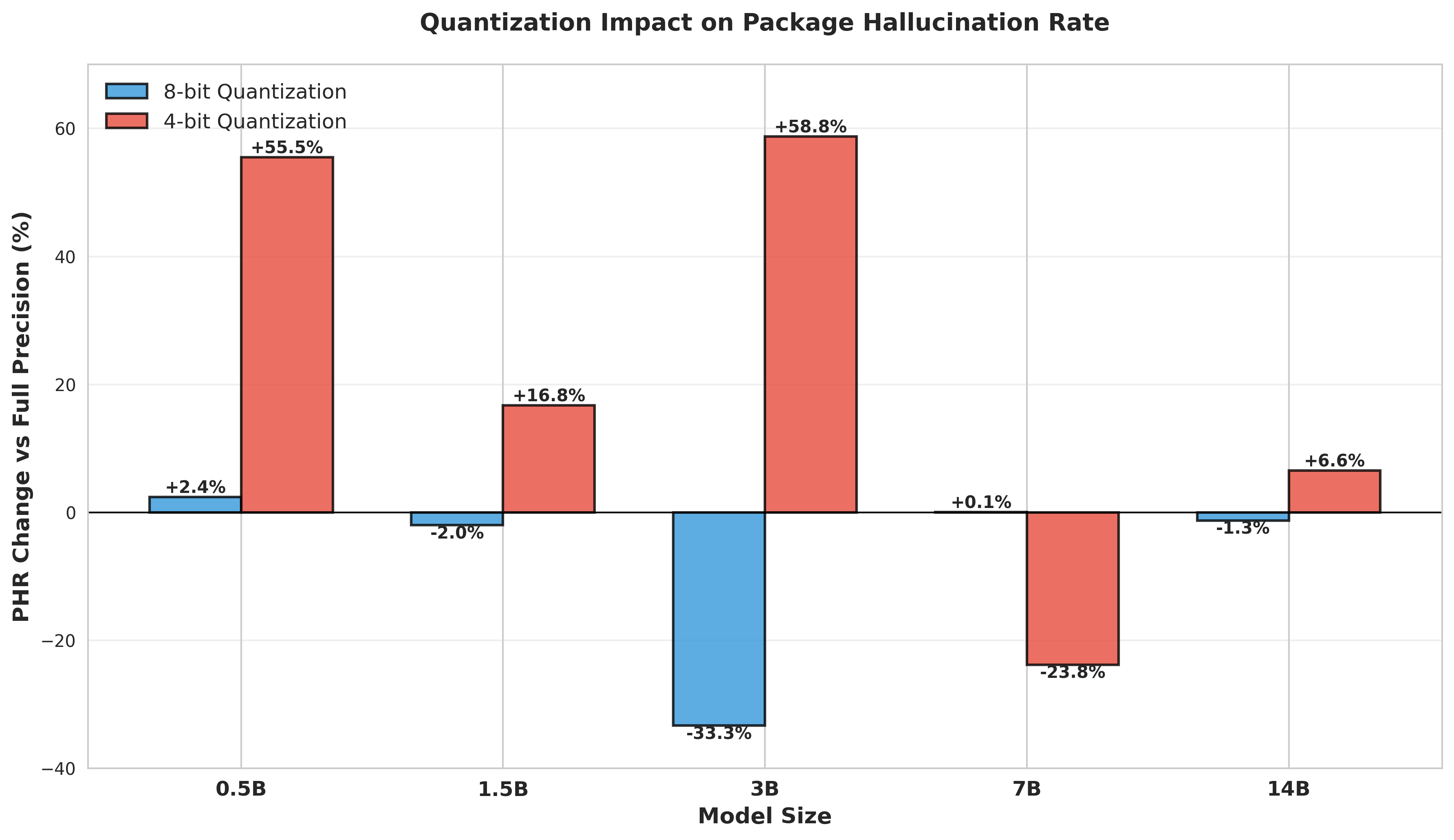}
        \caption{Quantization Degradation Impact}
    \end{subfigure}
    \hfill
    \begin{subfigure}
        [b]{0.49\textwidth}
        \centering
        \includegraphics[width=\textwidth]{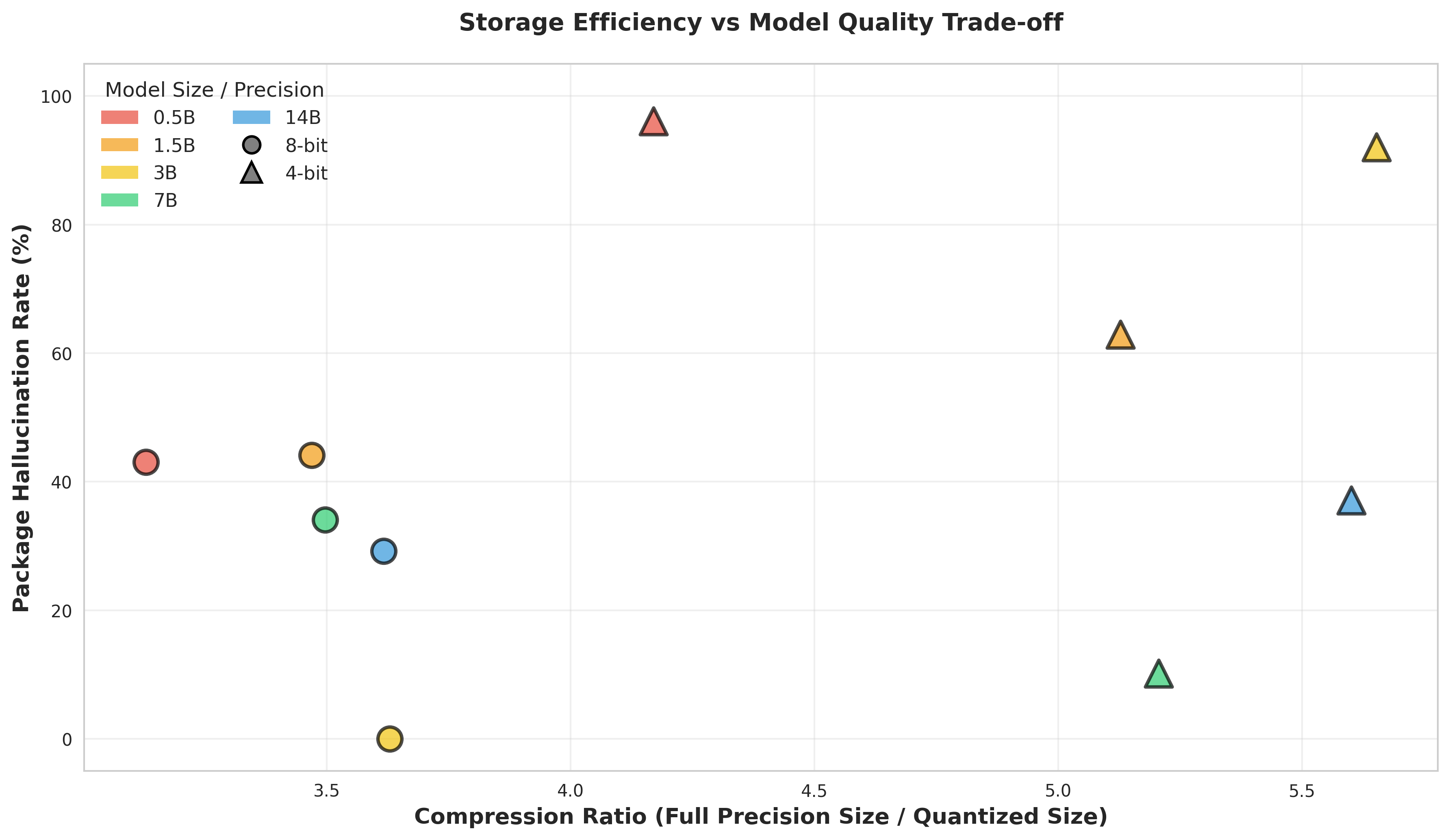}
        \caption{Storage Efficiency vs Model Quality Trade-off}
    \end{subfigure}
    \caption{Quantization Trade-offs: Impact Analysis and Storage Efficiency.}
    \label{fig:discussion}
\end{figure*}
Our analysis reveals critical trade-offs between model compression and package hallucination rates across different quantization levels. As shown in \autoref{fig:discussion},  8-bit quantization consistently achieves 3.1-3.6X compression with negligible quality impact (<2.5\% PHR change) across all model sizes except the 3B model. Specifically, the 1.5B, 7B, and 14B models demonstrate optimal compression characteristics, where 8-bit quantization either maintains or slightly improves hallucination rates while reducing storage requirements by ~70\%. However, 4-bit quantization presents a fundamentally different challenge: while achieving 4.2-5.7× compression, it causes substantial quality degradation in smaller models (0.5B: +55.5\%, 3B: +58.8\% PHR increase). Conversely, larger models (7B, 14B) tolerate 4-bit quantization better, with 7B showing improvement (-23.8\%) and 14B showing minimal degradation (+6.6\%). These findings suggest that aggressive quantization strategies should be model-size dependent. For practical deployment, 8-bit quantization emerges as the recommended approach for resource-constrained environments, providing substantial storage savings without compromising model reliability. The non-monotonic relationship between compression and quality—where more compression does not always correlate with worse performance—indicates that quantization effects are model-specific and warrant empirical evaluation rather than blanket application across different model architectures.


As we see from our results in Section~\ref{sec:results}, LLMs can hallucinate golang packages.
One way an attacker could take advantage of this is by observing hallucinated responses and creating new malicious packages that matches the hallucinated responses.
The largest attack vector is through \emph{github.com}.
As anyone can create any username and repository name on github, the attacker can create malicious golang packages through github.

Existing golang packages on github could also be targets for attackers.
An attacker can also observe what existing packages are recommended the most often by the LLM.
These frequently recommended packages could be targets of compromised.
The attacker can compromise the repository through approaches such as pull requests and or leaked tokens on github.
Both approaches could allow an attacker to commit malicious code to the repository.

A compromised or malicious golang package can result in a software supply chain attack.
Software supply chain attacks happen when risks in software components results in cascading effects to the downstream.
Similarly, the vulnerabilities or malicious code in a golang package could propagate to all applications that include the package.

Another unexplored area is poisoning the LLM training data and guiding the LLM to respond with attacker created golang packages.
LLMs learn from their training data, which often include information on the internet.
An attacker could take advantage of this and distribute poisoned training data on the web.
The attacker could first create malicious go packages, then distribute information that would poison the LLM's training data.
These could include distributing go commands that install malicious packages the attacker created or distributing prompts that could guide a LLM to perform the attacker's intentions.

  \section{Threats To Validity and Future Improvements}
\label{sec:threatsToValidity}

Due to time constraints for this project, we evaluated the package strings using http requests for simplicity.
If a package string is reachable, then we determine that the package exists.
However, in some cases an unreachable url could still be downloaded as a golang package.
As our evaluation approach introduces threats to validity, for future work, we would like to improve our evaluation of package strings by executing \ttt{go} commands.
We can check the package by executing the \ttt{go~get} command, which downloads the package. 
This ensures that we are downloading packages and provides verification as to whether a golang package exists.

Other future improvements would include more rigorous testing of the LLMs.
For example, experimenting with different prompts and evaluating how the LLM responds differently for each.

  \section{Conclusion}
\label{sec:conclusion}
This study presents a comprehensive analysis of the impact of quantization on package hallucination and security vulnerabilities in \llmsforcode for shell command generation in the GoLang programming language. Through systematic evaluation across five model sizes and three precision levels, we have uncovered critical insights into the trade-offs between model efficiency and reliability. Our findings reveal that quantization consistently increases package hallucination rates, with 4-bit quantization causing particularly dramatic degradation in smaller models where hallucination rates can exceed 96\%. This relationship demonstrates a clear efficiency-reliability trade-off, where the computational and storage benefits of quantization come at the cost of significantly reduced accuracy in package generation.

Beyond increasing PHR, quantization also correlates with higher VPR that models do correctly generate, indicating that precision reduction not only makes models more likely to invent packages but also biases their valid outputs toward more vulnerable dependencies. This dual impact presents serious security implications for production code generation systems.

For practical deployment, 8-bit quantization provides a reasonable efficiency-accuracy balance for most applications, while 4-bit requires extreme caution and robust validation layers. Larger models show greater resilience to these effects, suggesting that model scale should inform quantization strategy. These findings highlight the critical need to weigh efficiency gains against code reliability and security in AI-assisted development tools.

  \bibliographystyle{ACM-Reference-Format}
  \bibliography{project/ref}

\end{document}